# Quasi-ballistic electron transport in as-produced and annealed multiwall carbon nanotubes

Hisashi Kajiura*, Anil Nandyala, and Alexey Bezryadin

*Department of Physics, University of Illinois at Urbana-Champaign, 1110 West Green Street, Urbana, Illinois 61801, USA*

* To whom correspondence should be addressed.

Fax: +81-46-226-2453. Tel: +81-46-226-3773

Electronic addresses: hisashi.kajiura@jp.sony.com

Postal address: Materials Laboratories, Sony Corporation, 4-16-1 Okata, Atsugi-city, Kanagawa, 243-0021, Japan

*Keywords:* A. Carbon nanotubes; B. Chemical vapor deposition, Heat treatment; C. Thermal analysis; D. Transport properties

The electronic properties of multiwall carbon nanotubes (MWNTs) [1] have attracted much attention because they could lead to nano-sized devices [2-6]. To obtain optimal performance, it is desirable to utilize MWNTs of the highest purity and the best quality. Using chemical vapor deposition (CVD), MWNTs can be produced in large quantities at a low cost [7]. CVD-produced MWNTs, however, contain a considerable amount of structural defects due to their low temperature synthesis [7,8]. Since these defects act as scattering centers in electron transport and thus limit the electronic mean free path (EMFP), high-temperature annealing should lengthen the EMFP. This is based on the previous experiments in which annealing eliminated structural defects in various carbon materials including nanotubes [8-10]. In this study we investigate how annealing affects the transport properties of CVD-grown MWNTs. To measure the EMFP of individual nanotubes, we submerged them into liquid mercury (Hg) and measured the variation in conductance [4]. We found that the EMFP is much longer in annealed nanotubes compared with that of as-produced ones. The annealed nanotubes show quasi-ballistic electronic transport with the EMFP reaching a few microns, even at room temperature.

We started with soot consisting of CVD-produced MWNTs (10-20 nm diameter, 1-5 µm length, >95% purity; NanoLab, Inc.). Scanning electron microscopy (SEM, JSM-6700F, JEOL) showed nanotubes protruding from the edges of the soot sample (Fig. 1A). The annealing was done in argon (>99.9999% purity) at $2700^{\circ}C$ for 30 min. The concentration of defects was estimated using thermogravimetric analysis (TGA, Pylis 1 TGA, Perkin-Elmer). Oxidation in carbon nanotubes begins at structural defects [11], and when nanotubes with many defects are subjected to an oxidative atmosphere, they begin to lose weight at a lower temperature than tubes with fewer defects. For the TGA measurements, the samples were dried at $105^{\circ}C$ for 1 h, and heated to $800^{\circ}C$ at $5^{\circ}C/min$ in dry air at 30 ml/min. The oxidation starting point, defined as the temperature at which the weight is reduced by 5%, increased from 360°C to 560°C with annealing (Fig. 1B). This indicates that certain types of defects are eliminated by annealing at 2700°C.

The conduction properties were measured in air at room temperature using a piezo-driven nanopositioning system (Fig. 2A) [12], which allowed gentle and reproducible contact between the sample, attached to the probe using silver paste, and the Hg counter electrode [4]. The mobile electrode ("probe") was attached to the piezo-positioner with a displacement range of 20 μm (17PAZ005, MELLES GRIOT). The Hg was positioned below the mobile electrode. To make electrical contact between the sample and Hg, the probe was driven cyclically up and down with a peak-to-peak amplitude of 2-10 μm and a frequency of 1 Hz. A potential of 180 mV was applied between the probe and Hg, and the current was measured as a function of the piezo-positioner displacement at a sampling rate > 1000 points/s using an analog-to-digital converter (NI 6120, National Instruments).

Figures 3A and B show conductance trace $G(x)$ for as-produced and annealed samples, respectively, normalized by the conductance quantum unit, $G_0 \equiv 2e^2/h = (12.9 \text{ k}\Omega)^{-1}$, where $e$ is the electronic charge and $h$ is Planck's constant. Here $x$ represents the piezo-positioner extension, with $x = 0$ corresponding to the point at which the tube-Hg contact is made. Thus $x$ measures the extent the nanotube segment is submerged into Hg (Fig. 2B). In the majority of the measurements, we observed a sequence of steps and plateaus in the $G(x)$ traces, each step corresponding to a new nanotube's making contact with Hg [4]. The plateau after each step shows that the conductance is almost independent of the length of the nanotube segment connecting the probe and the Hg electrodes. Thus, nanotubes act as quasi-ballistic conductors.

The difference between ballistic and diffusive conductors is that in diffusive conductors, electrons have many collisions with many scattering centers, while in ballistic conductors they can propagate through the sample with few collisions. A ballistic regime is realized if a conductor is not too large, has very few defects, and the temperature is sufficiently low. If a diffusive wire is connected to two bulk electrodes, the resistance of the system is proportional to the length of the wire. Contrary, in the case of a one-dimensional ballistic wire, the net resistance is $(nG_0)^{-1}$, independent of the wire length. Here $n$ represents the number of conduction channels ($n = 2$ for metallic carbon nanotubes) [13]. Thus the plateaus on the $G(x)$ curves reflect quasi-ballistic transport in nanotubes at room temperature.

Although a series of steps and plateaus was observed in most of the measurements, the theoretically expected $2G_0$ conductance jumps [13] were not found with our samples. The step size of $1G_0$ [4] was also not observed in our experiments. The first step was typically in the range $0.1$-$0.3G_0$. A possible explanation is that the nanotube, which touches the Hg electrode, may not make a direct contact to the probe. Instead, it is connected to other tubes in the soot, which implies that the contact resistance between the probe and the measured tube is high and random [14]. We believe that this contact resistance is responsible for small size of the first conductance step. If the tube-Hg contact resistance significantly contributes to the total resistance, a conductance plot will show a rounded shape close to $x = 0$ [15]. In our case, in both the as-produced and annealed samples, no rounded shape was observed, suggesting that the contribution of the tube-Hg contact resistance to the total resistance is negligible. Thus we assume that the tube-Hg contact resistance does not depend on $x$ in our estimation. We also assume that the effects of thermal activation and doping by gas adsorption on nanotube conduction are negligible [15].

Using the first plateau in the $G(x)$ trace, we estimated the resistance per unit length ($\rho$) of the nanotubes. For this, $G(x)$ was converted into resistance as $R(x) = 1/G(x)$ (Fig. 3C). The total resistance ($R$) can be expressed as $R = R_c - \rho x$, where $R_c$ is the contact resistance [15]. The linear fit in Fig. 3C (annealed sample) results in $R_c$ = 125±0.3 kΩ and $\rho$ = 1.3±0.4 kΩ/μm. The resistance plot from Fig. 3A (as-produced sample) yields $R_c$ = 79±0.1 kΩ and $\rho$ = 9.7±0.2 kΩ/μm. The $R_c$, obtained from 102 traces (53 as-produced and 49 annealed samples), was typically in the range 30-150 kΩ. The $\rho$ of a carbon nanotube is related to the EMFP ($\ell$) as $\rho = (h/4e^2)(1/\ell)$ [5]. Using this equation, we plot histograms of EMFP obtained from measurements of the as-produced and the annealed samples (Fig. 4). It is noteworthy that for the as-produced samples, the peak of the EMFP distribution is found at ~0.3 μm, while the annealed samples has a peak at ~2.0 μm, which is about 7 times higher than that of as-produced ones. The lowest EMFP for the annealed samples was 0.6 μm, which is 20 times higher than that for as-produced samples (0.03 μm).

In summary, we measured the electrical resistance of nanotubes by submerging them into Hg in order to characterize their transport properties. We tested CVD-produced MWNTs before and after high-temperature annealing. The results show that annealing considerably reduces the amount of defects and leads to a significant increase in the electronic mean free path, which reaches a few microns at room temperature.

List of captions

Figure 1    (A) SEM image showing MWNTs protruding from the edges of the soot. (B) TGA profiles for as-produced and annealed samples.

Figure 2    Schematic diagrams of (A) experimental setup and (B) tube-Hg contact. Here $x$ is the length of the nanotubes segment submerged into Hg.

Figure 3    Conductance traces for (A) as-produced and (B) annealed samples, normalized by $G_0 = 2e^2/h$, and (C) the resistance trace $R(x) = 1/G(x)$ (the first step of B) as a function of piezo-positioner displacement ($x$), measured as the tube was pushed into Hg. The straight line in (C) is the fit given by $R = C - \rho x$. Here C is one fitting parameter and $C \approx R_c$, and $\rho$ is the other fitting parameter representing the resistance per unit length of the nanotube.

Figure 4    Histograms of the electronic mean free path (EMFP) for (A) as-produced and (B) annealed samples.

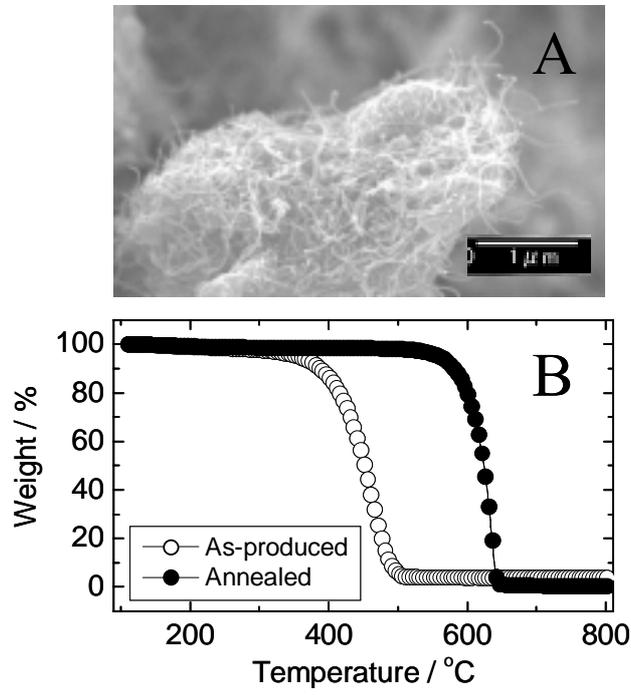

Figure 1 Kajiura *et al*.

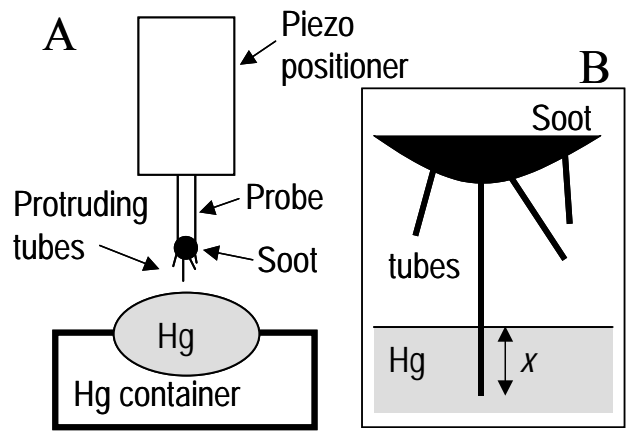

Figure 2 Kajiura *et al.*

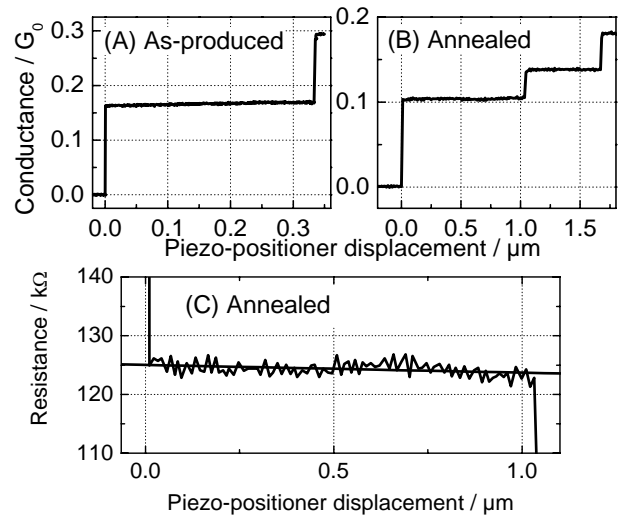

Figure 3 Kajiura *et al.*

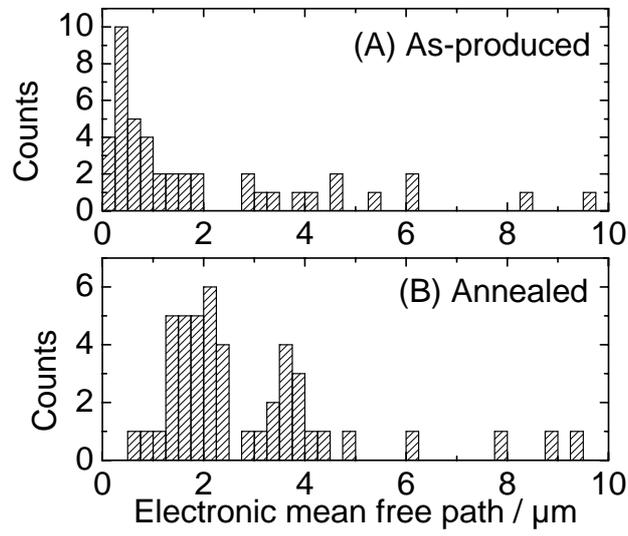

Figure 4 Kajiura *et al.*